\documentclass[a4paper,11pt]{article}
\usepackage[english]{babel}
\usepackage[latin3]{inputenc}
\usepackage[T1]{fontenc}
\usepackage{textcomp}
\usepackage{epsfig}
\usepackage{latexsym}
\usepackage{graphicx}
\usepackage{amsfonts,amssymb,bm}
\usepackage{color}
\newcommand{\D}{{\rm d}}
\newcommand{\sign}{{\rm sign\,}}

\newcounter{llista}

\begin{document}
\title{Einstein's accelerated reference systems \\
and Fermi-Walker coordinates}
\author{Josep Llosa\\
{\small Departament de Física Quàntica i Astrofísica, Institut de Ciències del Cosmos} \\ Universitat de Barcelona}

\maketitle

\begin{abstract}
We show that the uniformly accelerated reference systems proposed by Einstein when introducing acceleration in the theory of relativity are what is known at present as Fermi-Walker coordinate systems. We then consider more general accelerated motions and, on the one hand we obtain Thomas precession and, on the other, we prove that the only accelerated reference systems that at any time admit an instantaneously comoving inertial system belong necessarily to the Fermi-Walker class.

\end{abstract}

\section{Introduction\label{S0}}
In the final parts of ``On the relativity principle and the conclusions drawn from it'' \cite{Einstein1907}\footnote{I actually follow the English version of the article as presented in \cite{collected} } Einstein started the endeavor that would culminate in his relativistic theory of gravitation. Particularly he poses the question ``Is it conceivable that the principle of relativity also applies to systems that are accelerated relative to each other?'' and, in section \S 18 he tries to extend the framework of his Theory of Relativity by setting up a theory of uniformly accelerated reference systems. With the help of this and the Equivalence principle advanced in section \S 17, he could derive his first gravitational redshift formula and pointed to Sun's spectral lines for a test.

The aim of the present work is not a critical reading from the vantage point of the knowledge we now have; it would be both unfair and senseless. Nevertheless it might be worth to translate the main notions in the above mentioned fragment into today language, so benefiting from the possibilities of spacetime mathematical framework that Minkowski \cite{Minkowski1908} would set up just a few months after the issue of \cite{Einstein1907}. This would be helpful for a present time reader to better understand that passage in which Einstein tackles the problem of accelerated reference frames in relativity. It will also help to appraise the sharp insight achieved by Einstein in spite of the shortage of theoretical means and the lack of the mathematical framework of Minkowski spacetime which is the most suitable for that task.

We shall see how the accelerated reference systems advanced by Einstein are no other than Fermi-Walker coordinate systems (without rotation) whose development, with the suitable mathematical tools, would start more than one decade later \cite{Fermi-Walker}. 
Our work also intends to throw some light on alternative ways to extend the class of inertial systems of special relativity to abide accelerated systems, what are the postulates, what are the (often tacit) underlying assumptions and what apparently harmless suppositions are inconsistent.

We first list and label a transliteration of the assumptions on which Einstein's construction is based:
\begin{list}
{\bf A\arabic{llista}}{\usecounter{llista}}
\item ``We first consider a body whose individual material points, at a given time $t$ of the nonaccelerated reference system $\mathcal{S}$, posses no velocity relative to $\mathcal{S}$, but a certain acceleration. \,\ldots \, we do not have to assume that the acceleration has any influence on the shape of the body.''
\item `` \ldots a reference system $\Sigma$ that is uniformly accelerated relative to the nonaccelerated system $\mathcal{S}$ in the direction of the latter's $X$-axis''. 
\item `` \ldots  and the axes of $\Sigma$ shall be perpetually parallel to those of $\mathcal{S}$''
\item ``At any moment there exists a nonaccelerated reference system $\mathcal{S}^\prime$ whose coordinate axes coincide with the coordinate axes of $\Sigma$ at the moment in question (at a given time $t^\prime$ of $\mathcal{S}^\prime$).''
\item `` If the coordinates of a point event occurring at this time $t^\prime$ are $\xi$, $\eta$, $\zeta$ with respect to $\Sigma$, we will have
$$ x^\prime = \xi\,, \qquad  y^\prime = \eta\,, \qquad z^\prime = \zeta\,$$
because in accordance with what we said above \ldots ''
\item `` \ldots the clocks of $\Sigma $ are set at time $t^\prime$ of $\mathcal{S}^\prime$ such that their readings at that moment equal $t^\prime$.''
\item `` \ldots a specific effect of {\em acceleration} on the rate of the clocks of $\Sigma$ need not be taken into account.''
\item `` \ldots the clocks of $\Sigma$ are adjusted  \ldots at that time $t=0$ of $\mathcal{S}$ at which $\Sigma$ is instantaneously at rest relative to $\mathcal{S}$. The totality of readings of the clocks $\Sigma$ adjusted in this way is called the \guillemotleft local time\guillemotright\, $\sigma$ of the system $\Sigma$.''
\item `` \ldots two point events occurring at different points of $\Sigma$ are not simultaneous when their local times $\sigma$ are equal.''
\item `` \ldots the \guillemotleft time\guillemotright\, of $\Sigma$ [is] the totality of those readings of the clock situated at the coordinate origin of $\Sigma$ which are [ \ldots ] simultaneous with the events which are to be temporarily evaluated.'' 
\end{list}
The labels {\bf A1}, {\bf A2}, \ldots will serve to indicate which assumption in the above list is being invoked at each moment along the discussion that follows.

It is worth to remark here that {\bf A6} and {\bf A10} are somehow equivalent and define $\Sigma$-simultaneity as determined by equal time $t^\prime$, i. e. by $\mathcal{S}^\prime$-simultaneity. 
On their turn, {\bf A7}, {\bf A8} and {\bf A9} refer to the behaviour of local cloks that are stationary in $\Sigma$. Assumption  {\bf A7} is what is called {\em clock hypothesis} ---see for instance \cite{Rindler77}--- which implies that 
in today's words this \guillemotleft local time\guillemotright amounts to \guillemotleft proper time\guillemotright.
{\bf A8} establishes how the swarm of $\Sigma$ local clocks are set at zero and {\bf A9} is not actually an assumption but a consequence of {\bf A7} and {\bf A8} together with previous assumptions, as it will be seen in Section \ref{S3.1}.

Furthermore, {\bf A4} does not explicitly state that $\Sigma$ is instantaneously at rest relative to $\mathcal{S}^\prime$. However, the ``said above'' in {\bf A5} refers to applying {\bf A1} to the relation between the systems $\Sigma$ and $\mathcal{S}^\prime$ at the moment $t^\prime$, which amounts to tacitly state that both systems are (instantaneously) at rest with respect to each other, that is
\begin{description}
\item[{A4'}] {At any moment there exists a nonaccelerated reference system 
$\mathcal{S}^\prime$ whose coordinate axes coincide with the coordinate axes of $\Sigma$ at the moment in question (at a given time $t^\prime$ of $\mathcal{S}^\prime$) and such that $\Sigma$ is instantaneouly at rest with respect to $\mathcal{S}^\prime\,$. }
\end{description}
In what follows we shall rather use assumption {\bf A4'} instead of {\bf A4}.

In Section \ref{S2} we present a summary of preliminary notions, such as Fermi-Walker transport and Fermi-Walker coordinates. Then in section \ref{S3} we translate Einstein's reasoning into today's language and find that his uniformly accelerated systems of coordinates are actually Fermi-Walker coordinates \cite{Fermi-Walker} for one-directional motion. 

Section \ref{S4} analyzes the relative importance of the assumptions in the list for the final result and in what cases they might incur in  conflict or inconsistency, e. g. if the one-directionality condition in {\bf A2} is dropped, then the $\Sigma$ axes cannot be perpetually parallel to those of $\mathcal{S}$, provided that {\bf A4'} and {\bf A5} are  mantained. 
We also prove in Section \ref{S4} that the only systems of coordinates compatible with assumptions {\bf A1}, {\bf A4'} and {\bf A5} are Fermi-Walker systems for an arbitrary proper acceleration of the origin of coordinates.

\section{Preliminary notions: Fermi-Walker coordinates \label{S2} }
Roman indices $a, b, c, \ldots $ run from 1 to 4 whereas indices $i, j, k, \ldots$ run from 1 to 3; the convention of summation over repeated indices is adopted everywhere and, for the sake of simplicity, we take $c=1$ and $x^4 = t$. Tensor indices are raised/lowered with Minkowski metrics $\eta_{ab} = {\rm diag}(1\,1\,1\,-1)$\,, so that $M^j = M_j$ and $M^4 = -M_4\,$.

Let  $\,z^a(\tau) \,$ be a proper time parametrized timelike worldline in ordinary Minkowski spacetime, which we shall take as the {\em space origin}, $a=1 \ldots 4\,$, $u^a = \dot{z}^a(\tau)\,$ is the unit velocity vector and $a^a = \ddot{z}(\tau) \,$ the proper acceleration. 

A 4-vector $\,w^a(\tau)\,$ is said Fermi-Walker (FW) transported \cite{Synge1} along $z^a(\tau)\,$ if 
\begin{equation}  \label{a1}
\frac{\D w^a}{ \D \tau} = \Omega^a_{\;\;b}\, w^b \,,\qquad \qquad {\rm with} \qquad \qquad \Omega^a_{\;\;b} = u^a a_b - u_b a^a 
\end{equation}

If $\,z^a(\tau) \,$ is a straight line ($a^a=0 \,$ and uniform rectilinear motion), Fermi-Walker transport coincides with parallel transport and the equation above reduces to $w^a = $constant. 
It is obvious that $u^a(\tau) \,$ and $a^b(\tau) \,$ are  FW transported along $\,z^a(\tau) \,$ but, generally, they are not parallel transported. 

Let us now consider an orthonormal tetrad $e^a_{(c)}(\tau) \,$, $c = 1 \ldots 4\,$, that is FW transported along $z^a(\tau) \,$ and such that $e^a_{(4)} = u^a \,$. The FW transport law (\ref{a1}) can be written in two equivalent forms:
\begin{equation}  \label{a2}
\frac{\D e^a_{(c)}}{ \D \tau} = \Omega^a_{\;\;b}\, e^b_{(c)}
\end{equation}
or equivalently
\begin{equation}  \label{a2b}
\frac{\D e^a_{(c)}}{ \D \tau} = \hat\Omega^d_{\;\;c}\, e^a_{(d)} \,, \qquad {\rm with} \qquad \hat\Omega^4_{\;\;i} =\hat\Omega^i_{\;\;4} = \hat{a}_i \,, \qquad 
\hat\Omega^i_{\;\;j} = 0
\end{equation}
where $0_3\,$ is the null square $\, 3 \times 3\,$ matrix and $\,\hat{a}_j = a_b  e^b_{(j)} \,$. Notice that 
$\,\hat\Omega_{dc} = \Omega_{ab} e^a_{(d)} e^b_{(c)}\,$. Both expressions, (\ref{a2}) and (\ref{a2b}) are equivalent; using a simile from rigid body kinematics, the first one corresponds to {\em spacetime axes} whereas the second one belongs to {\em body axes}.  
 
For a given point in spacetime with inertial coordinates $\,x^a$, the Fermi-Walker coordinates \cite{Synge1}, \cite{Gourgoulon} with space origin on $z^a(\tau)$ are:
\begin{description}
\item[The time] $\tau(x^b)$,  given as an implicit function by
\begin{equation}  \label{a3}
  \left[x^a - z^a(\tau)\right]\,u_a(\tau) = 0
\end{equation}
\item[The space coordinates], defined by
\begin{equation}  \label{a4}
 \xi^i = \left[x_a - z_a(\tau(x))\right]\,e^a_{(i)}(\tau(x))
\end{equation}
\end{description}

In order to derive the inverse transformation, we include that (\ref{a3}) implies that $x^a - z^a(\tau)$ is orthogonal to $\,u^a(\tau)=e^a_{(4)}(\tau)\,$, therefore  $x^a - z^a(\tau)$ is a linear combination of the spatial triad $\,e^a_{(j)}(\tau)\,$, $j=1,2,3$, which using (\ref{a4}) leads to $x^a - z^a(\tau) = \xi^j e^a_{(j)}(\tau)\,$. Hence, the inverse coordinate transformation $\varphi^a(\tau, \xi^1, \xi^2,\xi^3) \,$ is
\begin{equation}  \label{a6a}
\varphi^a(\tau,\vec\xi) = z^a(\tau) + \xi^j e^a_{(j)}(\tau)
\end{equation}

By differentiating this relation, it easily follows that
\begin{equation}  \label{a5a}
\D x^a = u^a\,\left[1 +\vec{\xi}\cdot\vec{a}(\tau)\right]\, \D\tau + e^a_{(i)}\, \D\xi^i
\end{equation}
where $a_ i(\tau) = a_b(\tau)\,e^b_{(i)}(\tau)$ and, for the sake of brevity, the ordinary vector notation in three dimensions has been adopted, namely $\,\vec{\xi}\cdot\vec{a} = \xi^1 a_1 + \xi^2 a_2 + \xi^3 a_3\,$. And the invariant interval, $\D s^2 = \eta_{ab}\D x^a \D x^b \,$, in FW coordinates becomes 
\begin{equation}  \label{a6}
\D s^2 = \D\vec{\xi}^2 - \left[1 + \vec{\xi}\cdot\vec{a}(\tau) \right]^2 \D\tau^2
\end{equation}

Imagine now a material body that is at rest in this FW coordinate system. The worldline of each material point will be $\xi^j = $constant in FW coordinates whereas, in inertial coordinates it will be given by equation (\ref{a6a}) for these precise constant values of $\xi^j$.

According to the invariant interval (\ref{a6}), the infinitesimal radar distance \cite{Landau},\cite{Born1909} between two close material points $\xi^j$ and $\xi^j +\D\xi^j$ in the FW coordinates is $\, \D l^2 = \D \vec\xi^2 \,$. The geometry of this body is therefore flat and rigid (independent of $\tau$, and FW coordinates are Cartesian coordinates.

Using the invariant interval formula with $\xi^j$ constant, we obtain that the proper time rate, ticked by a standard clock comoving with the material point $\xi^j$, is   
$$  \D\sigma =  \left[1 + \vec\xi\cdot\vec{a}(\tau) \right]\,\D\tau \,.$$
Therefore $\sigma$ and $\tau$ only coincide at the origin, $\xi^j = 0\,$. In general, $\sigma \neq \tau$ and usually the readings of proper time $\sigma$ by the stationary clocks at two different points $\vec\xi_1 \neq \vec\xi_2\,$ only will keep synchronized if $\,\left(\vec\xi_1 -\vec\xi_2\right)\cdot \vec a(\tau) = 0\,$, for all $\tau$ (this only has a solution if all directions $\vec a(\tau)\,$ keep in the same plane). Contrarily, since the invariant interval (\ref{a6})   does not contain cross terms, the coordinate time $\tau$ is locally synchronous, i. e. two events with the same $\tau$ on two close material points are simultaneous.

The factor $\,1 + \vec\xi\cdot\vec{a}(\tau) \,$ is also relevant in connexion with the domain of the FW coordinates, which does not embrace the whole Minkowski spacetime. Indeed, the procedure to obtain the FW coordinates of a point relies on solving the implicit function (\ref{a1}), which requires that the $\tau$-derivative of the left hand side does not vanish, that is $\,1 + \vec\xi\cdot\vec{a}(\tau) \neq 0\,$.

The proper velocity of the worldline $\vec\xi = $constant 
$$ U^a(\tau\,\vec\xi) = \frac{1}{\left| 1 + \vec\xi\cdot\vec{a}(\tau) \right|}\, \partial_\tau \varphi^a(\tau,\vec\xi) = u^a(\tau) \,\, \sign\left[1 + \vec\xi\cdot\vec{a}(\tau) \right] $$
To avoid time reversal we shall restrict the domain of the FW coordinates to the region $\,1 + \vec\xi\cdot\vec{a}(\tau) > 0\,$) and the hypersurface $\,1 + \vec\xi\cdot\vec{a}(\tau)  = 0\,$ or, in inertial coordinates, $\,1 + a_a(\tau)\,\left[x^a - z^a(\tau)\right] = 0\,$ is the horizon of the FW coordinate system.

As for the proper acceleration of the material point $\vec\xi$, we have
\begin{equation}  \label{a6aa}
A^b(\tau;\vec\xi) = \frac{\D u^b}{\D\tau}\,\frac{\D\tau}{\D\sigma} = \frac{a^b(\tau)}{1 + \vec\xi\cdot\vec{a}(\tau)}  \qquad\qquad {\rm and} \qquad \qquad a_j(\tau;\vec\xi) =  \frac{a_j(\tau)}{1 + \vec\xi\cdot\vec{a}(\tau) }\, ,
\end{equation}
which differs from one place to another. 

It is worth to mention here that Einstein's statement \cite{Einstein1913b}: \guillemotleft 
... {\em acceleration} possesses as little absolute physical meaning as {\em velocity} \guillemotright, 
does not hold {\em avant la lettre}. As a matter of fact, every place $\vec\xi$ in a FW reference space has a proper acceleration (\ref{a6aa}) which is measurable with an accelerometer. However, the laws of classical particle dynamics also hold in the accelerated reference frame provided that a field of {\em inertial} force $-a^b(\tau;\vec\xi)$ is included; the passive charge for this field being the inertial mass of the particle. It is only with this specification that two accelerated reference frames are equivalent from the dynamical (or even physical) viewpoint.

Also notice that, as local proper acceleration is different from place to place, there is not such a thing as {\em the acceleration} of a FW reference system.

\section{The accelerated reference system \label{S3}}
We proceed to translate the assumptions in the Introduction into modern spacetime language. 
The reference system $\mathcal{S}$ (assumption {\bf A1}), that we will hereafter denote $\mathcal{S}_0$, is inertial and its coordinates are $x^a$. In order to avoid a ``too intrinsic'' notation, all equations will be referred to these coordinates. 

Being at rest in the accelerated system $\Sigma$, the $\Sigma$-coordinates of every point in the material body $\Sigma$, $\vec{\xi} =(\xi^1,\xi^2,\xi^3)$, are constant. Written in $\mathcal{S}_0$ coordinates and proper time parametrized the worldline of such a material point is
$$ x^a=\varphi^a(\sigma,\vec{\xi})\,, \qquad \qquad a=1 \ldots 4 \,,$$
$\sigma$ is the proper time and is ticked by a local standard clock comoving with this material point. The ``local time'' defined in assumptions {\bf A7} and {\bf A8} corresponds to the readings of these standard comoving clocks provided that they have been set to zero when they were at rest with respect to $\mathcal{S}_0$, i. e. when $x^4=0$, hence
\begin{equation}  \label{e1}
 \varphi^4(0,\vec{\xi}) = 0
\end{equation}
and, since the body $\Sigma$ is (instantaneously) at rest relative to $\mathcal{S}_0$, it is not ``deformed'' ({\bf A1}, {\bf A5}) hence
\begin{equation}  \label{e2}
 \varphi^j(0,\vec{\xi}) = \xi^j \,, \qquad \qquad j=1 \ldots 3
\end{equation}

Let
\begin{equation}  \label{e3}
 z^a(\tau) = \varphi^a(\tau,\vec{0}) 
\end{equation}
be the (proper time parametrized) equation of the worldline of the origin of $\Sigma$.
According to {\bf A1} and {\bf A4'}, for every event $z^a(\tau)$ there is an inertial reference system $\mathcal{S}_\tau$ ---$\mathcal{S}^\prime$ in the original statement--- such that the body $\Sigma$ is instantaneously at rest. That is, at all events in the swarm of worldlines $\varphi^a(\sigma,\vec{\xi})$ which are $\mathcal{S}_\tau$-simultaneous with $z^a(\tau)$, the proper velocity is the same and it is also equal to the proper velocity of $\mathcal{S}_\tau$, i. e. $u^a= z^a(\tau)$. $\mathcal{S}_\tau$-simultaneity also implies that all these events lay in the spacetime hyperplane which is orthogonal to $u^a$ and contains the event $z^a(\tau)$.

Moreover, at this $\mathcal{S}_\tau$-instant:
\begin{itemize}
\item the axes of $\Sigma$ and $\mathcal{S}_\tau$ coincide ({\bf A4'}),
\item $\mathcal{S}_\tau$ moves at the velocity $v^j =  u^j/u^4$ relative to $\mathcal{S}_0$ ({\bf A1}).
\end{itemize}

\begin{figure}[htb]   \label{f1}
\begin{minipage}{7cm}
\begin{center}
\includegraphics[height=5cm]{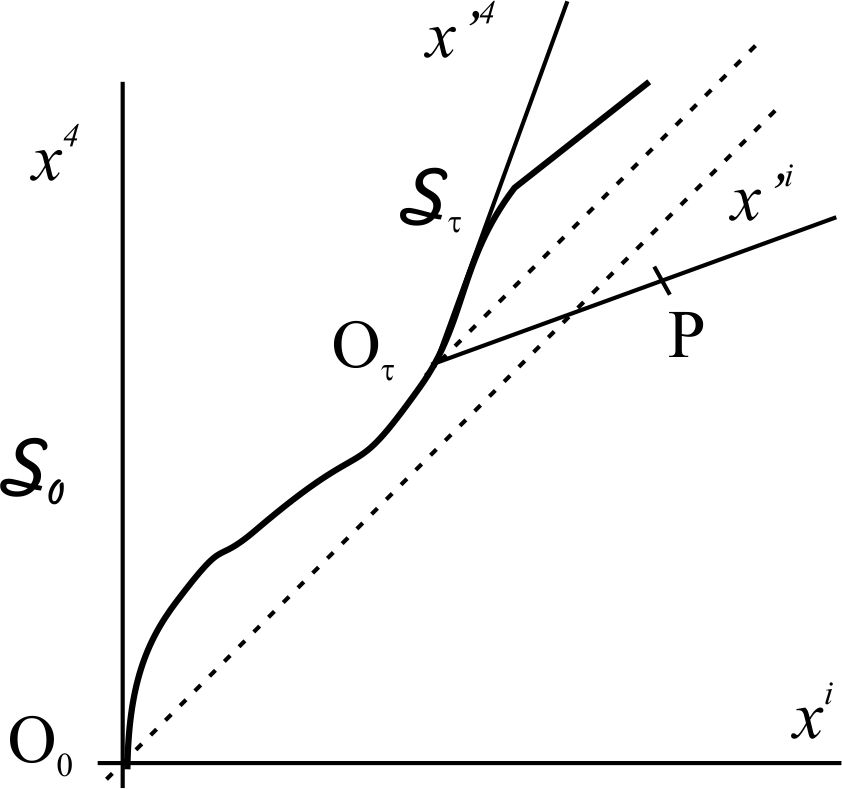}
\end{center}
\end{minipage}
\  \hspace*{.5em}  \
\begin{minipage}{7cm}
\begin{center}
\begin{tabular}{c|c|c|c}
\multicolumn{2}{c|}{Coordinates of:} & $P$ & $O_\tau$  \\
\hline
 $\mathcal{S}_0$ & \multicolumn{2}{|c|}{$x^a$} & $z^a(\tau)$ \\
\hline
$\mathcal{S}_\tau$ & \multicolumn{2}{|c|}{$x^{\prime b}$} & $ (0, \ldots \tau) $ \\
\hline
$\Sigma$ & $\xi^j = x^{\prime j}$ & $\tau = x^{\prime 4}$ & $ (0, \ldots \tau) $ \\
\hline
  & ({\bf A4'}) & ({\bf A6}) &   
\end{tabular}
\end{center}
\end{minipage}
\caption{}
\end{figure}

Consider the spacetime diagram in figure \ref{f1} where, for the sake of simplicity, the three space axes have been reduced to only one \\[1ex]

\bigskip
The worldline $z^a(\tau)$ of the origin of $\Sigma$ is depicted and two events, $O_0$ and $O_\tau$, are singled out. $\mathcal{S}_0$ and 
$\mathcal{S}_\tau$ are the inertial frames that, according to {\bf A1} and {\bf A4'}, are instantaneously at rest with respect to $\Sigma$.

Since both,  $\mathcal{S}_0$ and $\mathcal{S}_\tau$, are inertial frames, the coordinates $x^b$ and $x^{\prime a}$ of an event in each system are connected by a Poincaré transformation 
\begin{equation}  \label{e3a}
 x^a = \Lambda^a_{\;\;b} x^{\prime b} + s^a 
\end{equation}
where $\Lambda^a_{\;\;b}$ is a Lorentz matrix and  $s^a$ is a 4-vector. 

Particularly, the coordinates of the event $O_\tau$ in $\mathcal{S}_0$ are $z^a(\tau)$, whereas its $\mathcal{S}_\tau$-coordinates are
$x^{\prime i} = 0$ and $x^{\prime 4} = \tau$, which substituted in (\ref{e3a}) leads to
\begin{equation}  \label{e3aa}
 z^a(\tau) = \Lambda^a_{\;\;4} \tau + s^a 
\end{equation}
Furthermore, the $\mathcal{S}_\tau$-coordinates of an event $P$ which is $\mathcal{S}_\tau$-simultaneous with $O_\tau$ are $x^{\prime i} $ and $x^{\prime 4} = \tau$, where {\bf A6} has been included. Now since, by assumption {\bf A5}, $x^{\prime i} = \xi^{\prime i}$, the $\mathcal{S}_0$-coordinates of $P$ can be obtained using  equation (\ref{e3a})  and they are
$$ x^a = \Lambda^a_{\;\;j} \xi^{\prime j} + \Lambda^a_{\;\;4} \tau + s^a  $$
or, using (\ref{e3aa}),
\begin{equation}  \label{e3ab}
  x^a = \Lambda^a_{\;\;j} \xi^{\prime j} + z^a(\tau)
\end{equation}

On their turn the matrix elements $ \Lambda^a_{\;\;b}$ can also be seen as the components of the unit 4-vectors along the axes of the system $\mathcal{S}_\tau$ (and $\Sigma$) with respect to the orthonormal base of 4-vectors along the axes of $\mathcal{S}_0$, therefore
\begin{equation}  \label{e3ac}
 \tilde{e}^a_{(b)}(\tau) = \Lambda^a_{\;\;b}(\tau) \,, \qquad  b = 1\ldots 4 $$ 
and, as $u^a = \dot{z}^a(\tau)$ is the time axis of $\mathcal{S}_\tau$, we also have that
$$ u^a = \tilde{e}^a_{(4)}(\tau) = \Lambda^a_{\;\;4}(\tau) 
\end{equation} 

Equation (\ref{e3ab}) can thus be understood as
\begin{equation}  \label{e5a}
 (\xi^j,\tau) \longrightarrow  x^a = z^a(\tau) + \xi^j \tilde{e}^a_{(j)}(\tau) \,, \qquad 
\end{equation}
i. e. the coordinate transformation formula connecting  the systems $\Sigma $ and $\mathcal{S}_0$.

This equation recalls the formula (\ref{a6a}) for the inertial coordinates of an event in terms of its Fermi-Walker coordinates.  In order to prove that they were the same thing we should find out whether the tetrad of unit vectors $\,\tilde{e}^a_{(b)}(\tau) = \Lambda^a_{\;\;b}(\tau)\,$ is Fermi-Walker transported along the worldline $z^a(\tau)$.

As the spatial axes of $\mathcal{S}_0$, $\mathcal{S}_\tau$ and $\Sigma$ keep always parallel (recall {\bf A3} and {\bf A4'}), $\, \Lambda^a_{\;\;b} = \tilde{e}^a_{(b)} \,$ must be a boost matrix, hence it is completely determined by the relative velocity $v^j= u^j/u^4\,$, that is \cite{Rohrlich}
\begin{equation}  \label{e3b}
  \Lambda^i_{\;\;j} = \delta^i_j + \frac{u^i \,u^j}{1+u^4} \,, \qquad   \qquad 
	\Lambda^j_{\;\;4} =	\Lambda^4_{\;\;j} = u^j \,, \qquad  \qquad 	\Lambda^4_{\;\;4} = u^4 
\end{equation}
where we have included that
$$ \frac{v^i v^j}{v^2}\,(\gamma-1) = \frac{u^i u^j}{u^4 + 1} \,.$$

In a covariant form we can equivalently write
\begin{equation}  \label{e4}
\Lambda^a_{\;\;b}  = \delta^a_b - 2\,u^a n_b + \frac{(u^a + n^a) (u_b + n_b)}{1 - u^c n_c} \,, \qquad \qquad n^b = n_b =\delta^b_4
\end{equation}
This boost matrix depends only on the timelike 4-vectors $u^a$ and $n^a$, i. e. the time axes of the inertial frames it connects, and depends on $\tau$ only through $u^a= \dot{z}^a(\tau)\,$.

We must also realize that, being $\,\Lambda^a_{\;\;c}\,$ a Lorentz matrix,
\begin{equation}  \label{e6}
\Lambda^a_{\;\;c}\,\Lambda^{b c} = \eta^{ab}
\end{equation}
Differentiating the latter with respect to $\tau$, we have that $\,\dot{\Lambda}^a_{\;\;c}\,\Lambda^{b c} + \Lambda^{a c}\,\dot{\Lambda}^b_{\;\;c} = 0\,$, that is 
$$ W^{ ab} = \dot{\Lambda}^a_{\;\;c}\,\Lambda^{b c} \qquad \mbox{is skewsymmetric} $$
(a dot means derivative with respect to $\tau$) and, solving the latter for $\dot{\Lambda}^a_{\;\;c} \,$ and including (\ref{e6}), we arrive at the transport law
\begin{equation}  \label{e7}
\dot{\Lambda}^a_{\;\;c} = W^a_{\;\;b} \Lambda^b_{\;\; c} 
\end{equation}
or, equivalently, in terms of the tetrad $ \tilde{e}^a_{(c)}$, 
\begin{equation}  \label{e7a}
\frac{\D \tilde{e}^a_{(c)}}{\D \tau} = W^a_{\;\;b} \,\tilde{e}^b_{(c)} 
\end{equation}

On the other hand, comparing equation (\ref{e7}) with the result of a straight differentiation of  (\ref{e4}), we obtain
\begin{equation}  \label{e8}
W^{ab} = \frac{2}{1-n^c u_c}\, \left(u^{[a} + n^{[a}\right) \,a^{b]}
\end{equation}
where the square bracket means antisymmetrization, $\,a^b = \dot{u}^b(\tau) \,$ is the proper acceleration of the origin worldline $\,z^b(\tau)\,$ 
and we have included the fact that $n^b$ does not depend on $\tau\,$.

Now if we separate $n^b$ in its parallel and transverse parts with respect to $\,u^b(\tau)\,$
$$ n^a = - \left(n^c u_c \right)\,u^a + n_\perp^a\,, \qquad \qquad  n_\perp^a(\tau)\, u_a(\tau) = 0 \,,$$
equation (\ref{e8}) becomes
\begin{equation}  \label{e9a}
 W^{ab} = 2 \,u^{[a}\,a^{b]} + \frac{2}{1-n^c u_c}\, n_\perp^{[a} \,a^{b]}  
\end{equation}

Since by assumption {\bf A2} the system $\,\Sigma$ moves relative to $\mathcal{S}_0$ in the direction of the $X$-axis, the worldline $z^a(\tau)$ is contained in the spacetime plane $x^2 = x^3 =0$. This spacetime plane coincides with the one spanned by the 4-vectors $u^a(\tau)$ and $a^b(\tau)$, which remains constant. Moreover, since $n^a = u^a(0)$ and $u^b(\tau)  = n^b + \int_0^\tau a^b(\tau)\,\D\tau$, then $n^b$ is also coplanar with $u^b(\tau)$ and $a^b(\tau)$. Now, as both $n_\perp^b$ and $a^b(\tau)$ are orthogonal to $u^b(\tau)$ and coplanar with it, they must be parallel to each other and the second term in the right hand side of (\ref{e9a}) vanishes. So that, the skewsymmetric matrix is $\, W^{ab} = 2 \,u^{[a}\,a^{b]} \,$, the transport law (\ref{e7a}) reduces to 
\begin{equation}  \label{e10}
\frac{\D \tilde{e}^a_{(c)}}{\D \tau} = \Omega^a_{\;\;b} \,\tilde{e}^b_{(c)}  \qquad  \qquad {\rm with}  \qquad \qquad
 \Omega^a_{\;\;b} = u^a a_b - u_b a^a
\end{equation}
i. e. Fermi-Walker (FW) transport ---see eq. (\ref{a2}).

\paragraph{Summary:} Using assumptions {\bf A1}, {\bf A2}, {\bf A3}, {\bf A4'}, {\bf A5}, {\bf A6}, {\bf A7} and {\bf A8} in the list of nine assumptions proposed by Einstein \cite{Einstein1907}, we have obtained that, if an event has coordinates $\tau$, $\xi^1$, $\xi^2$, $\xi^3$ in the accelerated system $\Sigma$, then its inertial coordinates in $\mathcal{S}_0$ are given by equation (\ref{e3ab})
$$  x^a = z^a(\tau) + \xi^j \tilde{e}^a_{(j)}(\tau)  $$
where the orthonormal tetrad of 4-vectors $\,\left\{\tilde{e}^a_{(b)}(\tau) \right\}_{b=1\ldots4} \,$, with $\,\tilde{e}^a_{(4)} = u^a$, is Fermi-Walker transported along the worldline $\,z^a(\tau)\,$, just like Fermi-Walker coordinates discussed in Section \ref{S2}. Also, as a consequence of  {\bf A2}, the class of Einstein's accelerated systems are restricted to those whose origin of coordinates undergoes hyperbolic motion, i. e. rectilinear motion in space with constant proper acceleration.

\subsection{ Local time and [coordinate]  time \label{S3.1}}
In today's language the \guillemotleft local time\guillemotright \,introduced in {\bf A8} and {\bf A7} is the proper time parameter on the worldline of each material point in $\Sigma$, starting $\sigma=0$ at the event defined by (\ref{e1}). Therefore
\begin{equation}  \label{e11}
\varphi^a(\sigma,\vec{\xi}) = z^a(\tau) + \xi^j \tilde{e}^a_{(j)}(\tau)
\end{equation}
where $\tau$ is the coordinate \guillemotleft time \guillemotright in the accelerated system $\Sigma$, according to the assumption {\bf A10}. The relation $\tau = \tau(\sigma)$ can be obtained from the fact that the proper velocity $U^a = \partial_\sigma \varphi(\sigma,\vec{\xi}) \,$ is a unit vector. By differentiating equation (\ref{e11}) and including the transport law (\ref{e10}), we arrive at
$$ U^a = \frac{\D\tau}{\D\sigma}\, u^a\,\left(1+ a_b \tilde{e}^b_{(j)} \xi^j  \right)   $$
and the normalization condition  $U^a U_a = -1$ amounts to 
$$  \frac{\D\sigma}{\D\tau} = 1+ \xi^j \tilde{a}_j \,, \qquad {\rm where} \qquad \tilde{a}_j = a_b \tilde{e}^b_{(j)}     $$
where the fact that $u^a u_a = -1 $ has been included as well. 

In our case, by assumption {\bf A2}, $\Sigma $ is accelerated along the axis $\xi^1$ of the system $\mathcal{S}_\tau$; therefore the proper acceleration is 
 parallel to $\,\tilde{e}^b_{(1)} \,$:  
\begin{equation}  \label{e12}
 a^b = a\,\tilde{e}^b_{(1)} \,, \qquad {\rm and} \qquad \D\sigma = \left(1+ a \xi^1  \right) \,{\D\tau}
\end{equation}
which is equation (30) in ref. \cite{Einstein1907} provided that the suitable changes in notation are included\footnote{The $\sigma$ and $\tau$ occurring in ref. \cite{Einstein1907} are meant to be ``small''}.

The latter equation implies that the ratio between the flow rates of \guillemotleft local time \guillemotright, $\D\sigma$, and \guillemotleft time \guillemotright, $\D\tau$, depends on the $\Sigma$-coordinate $\xi^1$ and therefore two local clocks at places with different $\xi^1$, that were set at $\sigma=0$ when $\tau=0$, do not keep synchronized and, as  ---assumption {\bf A6}--- simultaneous events have the same coordinate time $\tau$, they will have different local time $\sigma$. We have thus derived {\bf A9} from other previous assumptions.  

It is worth to recall here that the main goal of the commented fragment in Einstein's paper is to derive the relation between \guillemotleft local time\guillemotright \,({\em proper time} in a modern language) and coordinate time. Equation (\ref{e12}) combined with the Equivalence principle eventually led Einstein to the redshift formula in a homogeneous gravitational field ---eq. (30a) in ref. \cite{Einstein1907}. 

\section{Reviewing the set of assumptions \label{S4}}
The assumptions listed in the Introduction are intertwined with each other and somewhat redundant. Let us elucidate how far can we go without invoking assumptions {\bf A2} and {\bf A3}. 

Recall that in Section 3 the formula of coordinate transformation (\ref{e5a}) 
$$   (\xi^j,\tau) \longrightarrow  x^a = z^a(\tau) + \xi^j e^a_{(j)}(\tau) \,,    $$
was derived on the basis of  {\bf A1}, {\bf A4'}, {\bf A5} and {\bf A6} only; $\tau$ and $\xi^j$ are the $\Sigma$-coordinates of the event $x^a$ and are also the $\mathcal{S}_\tau$-coordinates of this event. Besides $\mathcal{S}_\tau$ is an inertial system that is instantaneously comoving with $\Sigma$ at the instant $\tau$ (in the $\mathcal{S}_\tau$ clocks); $\,e^a_{(j)}(\tau) \,$, $j = 1\ldots 3$, are the space axes of both $\mathcal{S}_\tau$ and $\Sigma$, and $e^a_{(4)}(\tau) = u^a(\tau)$ is the proper velocity of $\mathcal{S}_\tau$ with respect to $\mathcal{S}_0$. 

Assumption {\bf A3} ---perpetually parallel axes--- was then invoked to include that the matrix $\,e^a_{(b)}\,$ is a Lorentz boost. In the general, since we dispense with {\bf A2} and {\bf A3},  $\,e^a_{(b)}=L^a_{\;\;b} \,$ is a general Lorentz matrix, which can be decomposed as the product 
\begin{equation}  \label{e13}
L^a_{\;\;b} = B^a_{\;\;c} R^c_{\;\;b} 
\end{equation}
where $\,B^a_{\;\;c}\,$ is a boost matrix and $\, R^c_{\;\;b}\,$ is a space rotation matrix, i. e.
$$ R^4_{\;\;j} =  R^j_{\;\;4} = 0 \,, \qquad \qquad 
 R^4_{\;\;4} = 1  \qquad \quad $$
and $\, R^i_{\;\;j}\,$ is a 3-space rotation matrix.

As in Section \ref{S3}, the Lorentz matrix satisfies that $\,L^a_{\;\;c}L^{bc} = \eta^{ab}\,$, which implies that
\begin{equation}  \label{e14}
\dot{L}^a_{\;\;b} = A^a_{\;\;c} L^c_{\;\;b} \,, 
\end{equation}
where $\,A^{ab}\,$ is a skewsymmetric matrix that, given the timelike unit 4-vector $\,u^a\,$, can be split as
\begin{equation}  \label{e15}
A^{ab} = u^a q^b - u^b q^a +\frac12\,\varepsilon^{abcd} u_c p_d
\end{equation}
where 
\begin{equation}  \label{e15b}
q^a = A^{ab} u_b \qquad {\rm and}  \qquad p_d = \varepsilon_{abcd} A^{ab} u^c
\end{equation}
are orthogonal to $u^a$ and $\,\varepsilon^{abcd}\,$ is the 4-dimensional Levi-Civita symbol ($\,\varepsilon^{1234}=1\,$).

The worldline of the material point whose $\Sigma$-coordinates are $\,\xi^j\,$ is
$$ \varphi^a(\tau,\xi^k) = z^a(\tau) + \xi^j L^a_{\;\;j}(\tau) \,,    $$
where we have included that $\,e^a_{(j)} = L^a_{\;\;j}\,$. The parameter $\tau\,$ is not in general the proper time along the worldline of the material point but rather the $\Sigma$-time according to {\bf A10}.

The unit 4-vector parallel to $\partial_\tau \varphi^a(\tau,\xi^k)\,$, namely
\begin{equation}  \label{e16}
U^a(\tau,\xi^k) = \gamma\,\left[ u^a(\tau) + \xi^j \dot{\Lambda}^a_{\;\;j}(\tau) \right] \qquad 
{\rm with} \qquad \gamma = \left( 1 - 2\,\xi^j  \dot{\Lambda}^a_{\;\;j} u_a - \xi^j \xi^k  \dot{\Lambda}^a_{\;\;j} \dot{\Lambda}^b_{\;\;k} \eta_{ab} \right)^{-1/2} \,, 
\end{equation}
yields the proper velocity of the material point $\,\xi^k\,$ at the $\Sigma$-instant $\,\tau\,$. According to {\bf A10}, these events, for all $\,\xi^k\,$, are $\mathcal{S}_\tau$-simultaneous.

By assumption {\bf A4'}, all material points in $\Sigma$ are instantaneously at rest with respect to the inertial system $\mathcal{S}_\tau$ and, as the proper velocity of the latter with respect to $\mathcal{S}_0$ is $\,u^a(\tau)\,$, both unit vectors must be the same:
$$ U^a(\tau,\xi^k) = u^a(\tau) \,, \qquad \qquad \forall \,\xi^k $$

Combining now this with equation (\ref{e16}), we obtain that $\,\xi^j \dot{L}^a_{\;\;j}\,$ is parallel to $\, u^a  \,, \,\,  \forall \,\xi^j \,$, 
or  equivalently,
\begin{equation}  \label{e17}
\dot{L}^a_{\;\;j} = - u_b \dot{L}^b_{\;\;j}\, u^a = a_b {L}^b_{\;\;j}\, u^a
\end{equation}
which, including (\ref{e14}) and (\ref{e15}), leads to
\begin{equation}  \label{e18}
\left[u^a(q_b - a_b) - q^a u_b +\frac12 \,\varepsilon^a_{\;\;bcd} u^c p^d  \right]\, {L}^b_{\;\;j} = 0
\end{equation}

This equation can be split in two parts:
\begin{eqnarray}  \label{e19a}
\parallel \, u^a & \Rightarrow & (q_b - a_b)\,e^b_{(j)} = 0 \\[1ex]  \label{e19b}
\perp \, u^a & \Rightarrow & \varepsilon^a_{\;\;bcd} u^c p^d  \,e^b_{(j)}= 0
\end{eqnarray}
where we have included that $\,e^b_{(j)} = {L}^b_{\;\;j}\,$ and used that $\,u_b L^b_{\;\;j} = L_{b4} L^b_{\;\;j} = 0\,$.

Finally, as $\,e^b_{(j)} \,, \quad j = 1 \ldots 3\,$ is a base of the space of vectors orthogonal to $u^b$, it easily follows from (\ref{e19a}) and (\ref{e19b}) that
$$ q^b = a^b \qquad {\rm and} \qquad p^b = 0 $$
which substituted in equation (\ref{e15}) yields 
\begin{equation}  \label{e20a}
A^{ab} = u^a a^b - u^b a^a\,
\end{equation}
and the transport law (\ref{e14}) becomes
\begin{equation}  \label{e20}
\frac{\D e^b_{(a)} }{\D \tau} =  \left( u^b a_c - u_c a^b\right)\, e^c_{(a)}  \,,
\end{equation}
i. e. the Fermi-Walker transport law, and therefore the coordinates $(\xi^j,\tau)$ are Fermi-Walker coordinates based on the worldline of the origin of $\Sigma$. 

\subsection{A possible conflict between assumptions A3 and A4'. Thomas precession \label{S4.1}}
Let us now see that assumption {\bf A3} is inconsistent with {\bf A1} and {\bf A4'}, unless it is supplemented with the assumption that acceleration is one-directional, i. e. the acceleration of the origin of $\Sigma$ is always in the same direction. To this end, we shall prove that the triad of spatial axes of the instantaneously comoving inertial system $\mathcal{S}_\tau\,$, $\, e^b_{(j)}(\tau)\,, \quad j= 1\ldots 3\,$, rotates with respect to the triad of axes $\, \tilde{e}^b_{(k)}(\tau)\,, \quad k= 1\ldots 3\,$
of an inertial reference system such that: {\bf (a)} is at rest with respect to $\,\mathcal{S}_\tau\,$ and {\bf (b)} its spatial axes keep perpetually parallel to $\, e^b_{(j)}(0)\,, \quad j= 1\ldots 3\,$, the spatial axes of the inertial system $\mathcal{S}_0$.

As both spatial triads are orthonormal bases of the subspace orthogonal to $u^a$, they are connected by a rotation 
$$    {e}^b_{(j)}(\tau) = R^k_{\;\;j}(\tau) \tilde{e}^b_{(k)}(\tau) \,,  \qquad   j= 1\ldots 3   $$
As commented in Section 3, the components of the perpetually parallel tetrad $\,\tilde{e}^a_{(c)}(\tau) \,$ in the base $\,{e}^a_{(b)}(0)\,$ 
are the boost matrix $\,\Lambda^b_{\;\;c}(\tau)\,$, whereas the components of the $\mathcal{S}_\tau$ tetrad in the same base are $\, L^d_{\;\;b}(\tau) \,$. Therefore
\begin{equation}  \label{e23}
L^b_{\;\;j}(\tau) = \Lambda^b_{\;\;k}(\tau) \, R^k_{\;\;j}(\tau) \,, \qquad   j, k = 1 \ldots 3 \,, \qquad b = 1 \ldots 4
\end{equation}
where $\, {R}^k_{\;\;j}(\tau)\,$ is a space rotation in three dimensions, $\,R^k_{\;\;j}(0)=\delta^k_j\,$ and 
$\,  R^k_{\;\;j}(\tau) R_l^{\;\;j}(\tau) = \delta^k_l \,$.

Differentiating 
with respect to $\tau$ and including (\ref{e7}) and (\ref{e14}) we obtain
$$ A^b_{\;\;c} L^c_{\;\;j} = \Lambda^b_{\;\;k}\dot{R}^k_{\;\;j}  + W^b_{\;\;c} \Lambda^c_{\;\;k} R^k_{\;\;j}  $$
This can easily be rewritten as 
\begin{equation}  \label{e21a}
\left( A_{bc} - W_{b c} \right)\,\Lambda^b_{\;\;k} \Lambda^c_{\;\;i} = \dot{R}_{lj} R_i^{\;\;j}
\end{equation}
or, equivalently,
\begin{equation}  \label{e24}
 \dot{ R}_{lj} R_i^{\;\;j} = \frac2{1-n_d u^d}\, \,n_{\perp [b} a_{c]} \, \Lambda_b^{\;\;i}  \Lambda^c_{\;\;l}   \,,
\end{equation}
where (\ref{e9a}) and (\ref{e20a}) have been included.

Using now the components of $a_b$ and $n_b$ in the perpetually parallel base:
\begin{equation}  \label{e24a}
 \tilde{a}_j = a_b \Lambda^b_{\;\;j} \,, \qquad  \tilde{n}_j = n_b \Lambda^b_{\;\;j} =  n_{\perp b} \Lambda^b_{\;\;j} 
\end{equation}
equation (\ref{e24}) becomes:
\begin{equation}  \label{e24b}
 \dot{ R}_{lj} R_i^{\;\;j} = \frac2{1-n_d u^d}\, \,\tilde{n}_{[i} \tilde{a}_{l]} 
\end{equation}

As $\,{ R}^l_{\;\;j}(0) = \delta^l_j \,$, the spatial axes of the system $\Sigma$ remain always parallel to the axes of the inertial system $\mathcal{S}_0$ if, and only if, $\,\dot{ R}^l_{\;\;j} = 0 \,$, which amounts to $\,\tilde{a}_{l} \propto \tilde{n}_{l} = \tilde{n}_{\perp l} \,$ or $\, {a}_{b} \propto n_{\perp b} \,$; therefore the unit vector $\,\hat{a}^b = a^b/a\,$ is
$$ \hat{a}^b = \beta\,\left(  n^b + u_c n^c\, u^b \right) \qquad {\rm with } \qquad \beta =  \left[ \left(u_c n^c\right)^2 -1 \right]^{-1/2} $$
Nevertheless the absence of rotation does not imply any restriction on the magnitude of proper acceleration $\, a = \sqrt{a^b a_b}\,$.

It follows easily from the latter that the derivative $\, \dot{\hat{a}}^b \,$ lies in the plane spanned by $\,u^b\,$ and $ \hat{a}^b\,$ and, as the latter is a unit vector,  $\,\dot{\hat{a}}^b \hat{a}_b= 0 \,$, whence it follows that $\,\dot{\hat{a}}^b = \beta\, u^b \,$, where the factor $\beta = a\,$, as it can be easily determined from the derivative of $\,\hat{a}_b\hat{a}^b = 1\,$.

All this implies that the Fr\^enet-Serret system for the worldline $z^b(\tau)$ is
\begin{equation}  \label{e25}
 \dot{u}^b  =   a\, \hat{a}^b \,, \qquad  \dot{\hat{a}}^b  = a \,u^b 
\end{equation}
where $\,a(\tau)\,$ is undetermined, and the solution for the initial data $\,u^b(0) = n^b\,$ and $\,\hat{a}^b(0) = \hat{w}^b\,$ is
\begin{equation}  \label{e26}
u^b(\tau) = \cosh \zeta(\tau)\, n^b + \sinh\zeta(\tau)\, \hat{w}^b \,, \qquad  
\hat{a}^b(\tau) = \sinh \zeta(\tau)\, n^b + \cosh\zeta(\tau)\, \hat{w}^b  
\end{equation}
with $\dot\zeta = a\,$, which corresponds to a planar curve in spacetime, i. e. to rectilinear spatial motion. 

Notice that $\,\hat{w}^b = \hat{a}^b(0) \,$ must be orthogonal to $n^b = \delta^b_4 = u^b(0)\,$ and  therefore the space components $a^j$ of the acceleration keep parallel to the constant space direction $w^j$
     
We have thus proved that the spatial axes of the instantaneously comoving inertial system $\mathcal{S}_\tau$ can keep perpetually parallel to the axes of the inertial system $\mathcal{S}_0$ if, and only if, the origin of coordinates of $\Sigma$ moves in a constant direction as seen from $\mathcal{S}_0$, i. e. if its acceleration has a constant direction, although it is not necessarily constant in magnitude. Therefore, the assumption that the acceleration of $\Sigma$ has a constant direction is necessary for assumption {\bf A3} to be consistent.

If, on the contrary, the spatial direction of proper acceleration is variable (as seen from $\mathcal{S}_0$), then $a^c$ cannot keep parallel to $n_\perp^c$ and, by (\ref{e24b}), $\,\dot{ R}^l_{\;\;j} \neq 0\,$. Hence, according to the inertial system $\mathcal{S}_0$, the spatial axes $\,e^a_{(j)}\,$ ---that by {\bf A4'} coincide with those of $\mathcal{S}_\tau$--- undergo a rotational motion which is known as {\em Thomas precession} \cite{Thomas} ---see the Appendix for details. 

\section{Conclusion}                     
We have studied the uniformly accelerated reference systems introduced by Einstein \cite{Einstein1907} when attempting to enlarge his theory of relativity to abide non-inertial systems of reference, in what would be the starting point of his path to his general theory of relativity. We have analyzed the list of assumptions in which Einstein's construct is based and, using the vantage point provided by the spacetime formalism, we have proved that Einstein's uniformly accelerated systems are a particular instance of Fermi-Walker coordinate systems \cite{Fermi-Walker}, \cite{Synge1}, namely when the proper acceleration of the origin of coordinates is constant both in direction and magnitude.

We have then examined the set of assumptions in which the Einstein reasoning is based and have found that some of them are partially redundant. Particularly the assumption that at every moment it exists an instantaneously comoving inertial system  ---assumption {\bf A4`}--- is a very restrictive postulate which unavoidably leads to Fermi-Walker coordinate systems. It is actually so strong that, as proved in section \ref{S4.1}, it gets in conflict with the assumption of perpetual parallelism of spatial axes ({\bf A3}), unless the condition of one-directional proper acceleration is added ({\bf A3}), as Einstein's construct does by the way. We show that the possibility of conflict among both assumptions, namely the existence an instantaneously comoving reference system and that the spatial axes keep perpetually parallel, lies in what is known as Thomas precession \cite{Thomas}.

\section*{Acknowledgment}
Funding for this work was partially provided by the Spanish MINECO under MDM-2014-0369 of ICCUB (Unidad de Excelencia 'María de Maeztu') and by Ministerio de Economia y Competitividad and ERDF (project ref. FPA2016-77689-C2-2-R).
 
\section*{Appendix: Thomas precession}
To evaluate the precession angular velocity, we define the components of angular velocity in the perpetually parallel base as 
$$ \tilde{\omega}^i = \frac12\,\varepsilon^{ilk}  \frac{\D R_{kj}}{\D t} \,R_l^{\;\;j} \,,$$
where $\,t = z^4(\tau)\,$ is the time in the inertial system $\mathcal{S}_0$ and $\,\D t = u^4\, \D \tau = - u_b n^b\, \D \tau  \,$. Then  equation (\ref{e24b}) implies that
\begin{equation}  \label{e27}
 \tilde{\omega}^i = \frac1{u^4\left(1 + u^4\right)}\,\varepsilon^{ilk}\,\tilde{n}_{l} \tilde{a}_{k} 
\end{equation}

To determine the components $\,\tilde{n}_{l}\,$ and $\, \tilde{a}_{k} \,$ in terms of the components of proper velocity and acceleration, we use (\ref{e24a}) and (\ref{e4}) so obtaining
\begin{equation}  \label{e28}
\tilde{a}_j = a_j + \frac{a_4}{1+u^4}\,u_j \,, \qquad \qquad \tilde{n}_k = - u_k 
\end{equation}
where the fact that $\,a_b u^b = 0\,$ has been included.

Using now  the 3-velocity $v^i$ and the 3-acceleration $b^j$ of the origin of $\Sigma$ with respect to the inertial system $\mathcal{S}_0$, we have that
$$ u^j = \gamma \,v^j \,, \qquad u^4 = \gamma = \frac1{\sqrt{1-v^2}}\,, \qquad a^j = \dot\gamma\, v^j + \gamma^2\,b^j $$
(we take $c=1$) which, combined with (\ref{e28}) and substituted in (\ref{e27}), yields
$$  \tilde{\omega}^i = \frac{\gamma^2}{1 + \gamma}\,\varepsilon^{ilk}\,b_{l} v_{k}  $$
or, in an obvious 3-vector notation,
\begin{equation}  \label{e29}
 \vec{\omega}  = \frac{\gamma^2}{1 + \gamma)}\,\vec{b}\times\vec{v}
\end{equation}
where $ \vec{\omega} = \left(\tilde{\omega}^1,\tilde{\omega}^2,\tilde{\omega}^3\right)\,$ and so on.


\begin{thebibliography}{99}
\bibitem{Einstein1907}{Einstein A,  {\em Jahrb Rad Elektr} {\bf 4} (1907) 411}
\bibitem{collected}{{\em The Collected Papers of Albert Einstein. The Swiss Years: Writings, 1900-1909} English translation, Anna Beck, Trans., Peter Havas,Cons., Princeton University Press (1989)}
\bibitem{Minkowski1908}{Minkowski H, ``Space and Time'' in {\em The Principle of Relativity}, Dover (1952)}
\bibitem{Rindler77}{Rindler W {\em Essential relativity}, \S 2.13, Springer-Verlag (New York, 1977)}
\bibitem{Fermi-Walker}{Fermi E, {\em Atti Acad Naz Lincei Rend Cl Sci Fiz Mat Nat} {\bf 31} (1922) 51; Walker A G, {\em Proc Roy Soc Edinburgh} {\bf 52} (1932) 345; Misner C, Thorne K S and Wheeler J A, {\em  Gravitation}, p 170, Freeman (1973) }
\bibitem{Thomas}{Thomas L H, {\em Nature} {\bf 117} (1926) 514}
\bibitem{Gallison}{Galison P, {\em Einstein's clocks and Poincare's maps}, Norton (2003)}
\bibitem{Synge1}{Synge J L, {\em Relativity: the special theory}, North-Holland (1965) }
\bibitem{Gourgoulon}{Gourgoulhon E, {\em Special relativity in general frames}, Springer 2013 }
\bibitem{Landau}{Landau L and Lifschitz E, {\em The Classical Theory of Fields}, Pergamon (1985)  }
\bibitem{Born1909}{Born M, {\em Phys Zeitschr} {\bf 10} (1909) 814}
\bibitem{Einstein1913b}{Einstein A, {\em Phys Zeitschr} {\bf 14} (1913) 1249}
\bibitem{Rohrlich}{Rohrlich F, {\em Classical Charged Particles}, Addison-Wesley (1990)}

\end{thebibliography}
\end{document}